\documentclass[12pt,preprint]{aastex}

\newcommand{\lsim}{\raise0.3ex\hbox{$<$}\kern-0.75em{\lower0.65ex\hbox{$\sim$}}}
\newcommand{\gsim}{\raise0.3ex\hbox{$>$}\kern-0.75em{\lower0.65ex\hbox{$\sim$}}}
\newcommand{\propsim}{\raise0.3ex\hbox{$\propto$}\kern-0.75em{\lower0.65ex\hbox{$\sim$}}}

\begin{document}
    
\title{Circular polarisation in compact radio sources: Constraints on particle acceleration and  electron-positron pairs}

\author{C.-I. Bj\"ornsson\altaffilmark{1}}
\altaffiltext{1}{Department of Astronomy, AlbaNova University Center, Stockholm University, SE--106~91 Stockholm, Sweden.}
\email{bjornsson@astro.su.se}

\begin{abstract}
It is shown that the frequency distribution of the degree of circular polarisation for a homogeneous source is sensitive to the properties of the synchrotron emitting plasma. Most of the circular polarisation comes from the region around the turn-over frequency, where the synchrotron radiation becomes optically thick. However, nearly circular characteristic waves result in circular polarisation dominated by frequencies above the turn-over frequency, while for nearly linear characteristic waves it is dominated by frequencies below. Observations argue in favour of nearly circular characteristic waves. This implies a low energy cut-off in the electron distribution substantially below that corresponding to the turn-over frequency and, simultaneously, provides an upper limit to the fraction of electron-positron pairs.

\end{abstract}


\section{Introduction}
The standard model for compact radio sources is well established; energy generated close to the central black hole is streaming out in a jet-like structure \citep{b/z77, b/p82}. However, several of its tenets lack a firm physical underpinning. This includes the launching of the jet and the means of energy transport. A central question here is the material constituents of the plasma; i.e., whether it consists of electrons and protons or if there is a significant fraction of electron-positron pairs. Another issue is the process by which the synchrotron emitting particles get accelerated to relativistic energies. Diffusive shock acceleration, second order Fermi acceleration in a turbulent plasma or direct acceleration by the electric field generated in a reconnection process of the magnetic field have all been suggested to be the agent transferring energy to the radiating particles. Since the injection of particles into the acceleration process is likely to be different for these mechanisms, the low energy end of the particle distribution may be one way to distinguish between them.

Unfortunately, optical depth effects hide the low energy electrons from direct view. Likewise, the presence of positrons can not be addressed by flux measurements alone, since their emitted flux is identical to that of the electrons. In contrast, both of these aspects of the plasma have a direct bearing on the observed circular polarisation. Although this was realised early on \citep[e.g.,][]{pac73}, the observed low level of circular polarisation made it hard to draw any strong conclusions regarding the plasma properties. However, it was noted that although the circular polarisation varied more rapidly and with larger relative amplitude as compared to either the flux or linear polarisation, it only rarely changed sign \citep{w/d83,kom84}. This suggests the presence of a large scale magnetic field. On the other hand, several theoretical arguments lead one to expect an important role for turbulence; e.g., in the acceleration process \citep{bla18,zhd18}. The connection between the large and small scale properties of the magnetic field is another issue where observations of circular polarisation have the potential to contribute significantly.

The low level of observed circular polarisation narrowed down the type of questions that could be addressed. The increased accuracy with which circular polarisation can now be measured has opened up new possibilities \citep{mac00,ray00}. Although VLBI-observations are still challenging \citep{hom09}, spatially resolved studies of circular polarisation along the jet can be made \citep{war98,h/w04}. Furthermore, polarisation can be measured over a wide frequency range \citep{osu13} as well as at high frequencies \citep{agu17a}. In spite of this increase, both qualitatively and quantitatively, of the observations of circular polarisation, no clear understanding of its origin has emerged \citep{vit08,osu13}. Hence, its use as a plasma diagnostic is still limited. However, since the circular polarisation from the synchrotron emission process itself is quite simple, the observed rather complex behaviour suggests that transport effects may play a crucial role.

The transfer equation for polarised light in a homogeneous medium has an analytical solution \citep[e.g.,][]{j/o77}. However, it is quite involved and various approximations have been put forth to facilitate comparison to observations. The aim of the present paper is threefold: (1) To present an alternative form of the homogeneous solution, which uses the concept of characteristic waves. This is an extension of the discussion in \cite{bjo88}. Such a formulation makes possible a more transparent and physical description of the polarisation of the emergent radiation. Furthermore, it is argued that even when the characteristic waves couple, this form of the solution can account, at least qualitatively, for some of the effects of inhomogeneities. (2) The high frequency observations in the POLAMI survey \citep{thu17,agu17b} and the detailed, wide band observations of \cite{osu13} are discussed. It is shown how they can be given a relatively straightforward explanation; in particular, that both indicate the presence of nearly circular characteristic waves. (3) It is pointed out that some of the approximations in common use have limited validity and, hence, should be applied with care.  

The outline of the paper is as follows: A short introduction to the transfer equation and the main properties of characteristic waves are given in Section\,\ref{sect2}. The formulation of the transfer equation for a light ray in terms of characteristic waves and its solution are presented in Section\,\ref{sect3}. The results for a homogeneous source are discussed in Section\,\ref{sect4}, where special attention is given to the two limits of nearly circular and nearly linear characteristic waves. Observations are discussed in Section\,\ref{sect5} and the main points of the paper are summarised in Section\,\ref{sect6}.

\section{Polarisation transfer in a homogeneous medium} \label{sect2}

Plane waves are the solution to Maxwell's equations in a homogeneous medium. Hence, instead of considering the propagation of a general electromagnetic field, it is sufficient to restrict attention to its Fourier components, $ \exp [i(\mathbf{k}\cdot \mathbf{r} - \omega t)]$. Here,  $k = 2\pi/\lambda $ and $\omega = 2\pi \nu$, where $\lambda$ and $\nu$ are the wavelength and frequency, respectively. However, the plane waves do not correspond to the physically measurable electric  and magnetic fields $\mathbf{E}$ and $\mathbf{B}$ but rather to $\mathbf{D}$ and $\mathbf{H}$, whose relations to the physical fields are determined by the properties of the medium. In a plasma relevant for synchrotron sources it is usually assumed that the permeability plays a negligible role so that $\mathbf{H} = \mathbf{B}$ and that the influence of the plasma can be written in component form as $D_{\rm l} = (\delta_{\rm l,m} + (4\pi i/\omega) \sigma_{\rm l,m})E_{\rm m}$, where $\sigma_{\rm l,m}$ is the dielectric tensor. The indices (l,m) run over all three spatial coordinate (x,y,z), i.e., (l,m = x,y,z) and a repeated index implies summation. Since $\mathbf{D \cdot k} = 0$, one finds
\begin{equation}
	E_{\rm z} + \frac{4\pi i}{\omega} \sigma_{\rm z,m}E_{\rm m} = 0,
	\label{eq1}
\end{equation}
where $\mathbf{k}$ has been chosen to lie along the z-axis (see Figure\,\ref{fig1}). Since $| \sigma_{\rm l,m} | \sim c \kappa$, where $\kappa$ is the absorptivity of the plasma, the magnitude ratio between the longitudinal and transverse components of the electric field is $| E_{\rm z} / E_{\rm (x,y)}| \sim \kappa / k$. This is usually a very small number, i.e., the distance over which the radiation is absorbed is much larger than its wavelength. The weak anisotropy limit then corresponds to neglecting the longitudinal component of the electric field, in which case the the transfer equation can be written                         
\begin{equation}
	\left(\frac{\partial}{\partial t} - c\frac{\mathbf{k}}{k}\cdot \frac{\partial}{\partial \mathbf{r}}\right)
	\left(\frac{\partial}{\partial t} + c\frac{\mathbf{k}}{k}\cdot \frac{\partial}{\partial \mathbf{r}}\right)
	\mathbf{E} = -4\pi \frac{\partial}{\partial t}\mathbf{J},
	\label{eq2}	
\end{equation}
where $J_{\rm l} = \sigma_{\rm l,m} E_{\rm m}$ (l,m = x,y,) is the current induced by the electric field. 

Since the magnitude of the RHS of equation (\ref{eq2}) is $\sim \omega c\kappa |E|$, it is seen that $kc/\omega \sim 1 + \kappa/ k$. The first operator on the LHS can then be evaluated to give $-2i \omega$, while the second operator corresponds to the comoving derivative in a frame moving with velocity $c$; i.e., $c\,{\rm d}/{\rm d}s$. Hence, without loss of accuracy, Equation (\ref{eq2}) can be written as a first order differential equation
\begin{equation}
	\frac{\rm d}{{\rm d}s}\, E_{\rm l} = -\frac{2\pi}{c}\sigma_{\rm l,m}E_{\rm m},
	\label{eq3}
\end{equation}
where $s$ is the distance along a ray path.

The transfer equation in Equation (\ref{eq3}) can be rewritten as
\begin{equation}
	\frac{\rm d}{{\rm d}s}\, E_{\rm l}E_{\rm j}^* = -\frac{2\pi}{c}\left(\sigma_{\rm l,m}E_{\rm m}E_{\rm
	j}^* + \sigma_{\rm j,m}^* E_{\rm m}^* E_{\rm l}\right), 
	\label{eq4}
\end{equation}
where $(^*)$ denotes complex conjugate and (l,j,m = x,y).
The transfer equation is usually written in terms of the Stokes parameters defined as $I = |E_{\rm x}|^2 + |E_{\rm y}|^2$, $Q = |E_{\rm x}|^2 - |E_{\rm y}|^2$, and $U +iV = 2E_{\rm x}E_{\rm y}^*$. It is straightforward to show that Equation (\ref{eq4}) is equivalent to the standard formulation.

In the homogeneous case, Equation (\ref{eq4}) has an analytical solution \citep[e.g.,][]{j/o77}; however, it is rather complex. An alternative to the standard formulation is to start from Equation (\ref{eq3}). The Stokes parameters are then calculated only after the radiation has been transported through the medium rather than at the point of emission. Although the two methods are equivalent, as discussed briefly in \cite{bjo88}, the latter solution is helpful when trying to understand how the physical properties of the plasma affect the polarisation of the emerging radiation. The reason is that the standard solution is expressed in terms of the various plasma parameters (i.e., $\sigma_{\rm l,m}$), while the alternative solution is expressed in terms of the polarisation properties of the two characteristic waves ($K^{1,2}$) and their phase difference ($\Delta k$).

\subsection{Characteristic waves}\label{sect2a}

The dielectric tensor in Equation (\ref{eq3}) can be written
\[ \sigma_{\rm l,m} =
\frac{c\kappa}{4\pi}\left( \begin{array}{cc}
	1 & \Upsilon_{\rm V} - i\Upsilon_{\rm L}\\
	-\Upsilon_{\rm V} - i\Upsilon_{\rm L} & 1
\end{array} \right),
\]
where $\Upsilon_{\rm V} = \hat{\xi}_{\rm V} + i \xi_{\rm V}$ and $\Upsilon_{\rm L} = \hat{\xi}_{\rm U} + i \xi_{\rm U}$. The notation in this paper follows rather closely the one in \cite{j/o77}, except that in order to avoid confusion with the complex conjugate, ($\,\hat{}$\,) is used instead of ($^*$) to denote parameters accounting for the circular and linear birefringence of the plasma. All the $\xi$-parameters are normalised to the absorptivity; e.g., $\xi_{\rm V} = \kappa_{\rm V}/\kappa$ and $\xi_{\rm U} = \kappa_{\rm U}/\kappa$, where $ \kappa_{\rm V}$ and  $\kappa_{\rm U}$ are the absorption coefficients for the Stokes $V$ and $U$ parameters (see Appendix C). Furthermore, it proves convenient to use $\phi = -\pi /4$ (see Figure\,\ref{fig1}) instead of $\phi = 0$, as done in \cite{j/o77}, since this renders $K^1 = -K^2$ (see below). As a result, the roles played by the Stokes parameters $Q$ and $U$ interchange; e.g., synchrotron emission has no $Q$-component. Since $U+iV = 2E_{\rm x}E_{\rm y}^{*}$, this choice also brings forth the formal similarity between the linear and circular polarisation.

The eigenvalues obtained by diagonalising $\sigma_{\rm l,m}$ are given by
\begin{equation}
	\eta^{1,2} = \frac{c\kappa}{4\pi}\left(1 \mp i\sqrt{\Upsilon_{\rm V}^2 +\Upsilon_{\rm L}^2}\right)
	\label{eq5}
\end{equation}
and Equation (\ref{eq3}) can be solved directly for the two characteristic waves 
\begin{equation}
	\mathbf{E}^{1,2} = \mathbf{E}^{1,2}_{\rm o} \exp\left(-\frac{2\pi}{c}\eta^{1,2} s \right).
	\label{eq6}
\end{equation}
Their phase difference can be defined as
\begin{eqnarray}
	\Delta k & = & -\frac{2\pi}{c}(\eta^1 - \eta^2) \nonumber\\
	& = & i\kappa \sqrt{\Upsilon_{\rm V}^2 + \Upsilon_{\rm L}^2}.
	\label{eq7}
\end{eqnarray}
Likewise, the polarisation of the two characteristic waves, $K^{1,2} \equiv E_{\rm y}^{1,2}/E_{\rm x}^{1,2}$, are obtained as
\begin{eqnarray}
	K^{1,2} & = & \mp \frac{\delta k}{\Upsilon_{\rm V} - i\Upsilon_{\rm L}} \nonumber\\
	& = & \pm \sqrt{\frac{1-\rho}{1+\rho}},
	\label{eq8}
\end{eqnarray}
where $\delta k = \Delta k/\kappa$ is the normalised phase difference and
\begin{eqnarray}
	\rho & \equiv & i\frac{\Upsilon_{\rm V}}{\Upsilon_{\rm L}} \nonumber\\
	& = & \frac{\hat{\xi}_{\rm V} \xi_{\rm U} - \hat{\xi}_{\rm U} \xi_{\rm V} + i(\xi_{\rm V} \xi_{\rm U}
	+\hat{\xi}_{\rm V} \hat{\xi}_{\rm U})}
	{\hat{\xi}_{\rm U}^2 + \xi_{\rm U}^2}.
	\label{eq9}
\end{eqnarray}
It should be noted that there is a sign ambiguity in Equations (\ref{eq7}) and (\ref{eq8}) when evaluating the square root. It is shown below that this sign always enters in the product of $K^{1,2}$ and $\Delta k$. Hence, the choice is physically unimportant as long as the same sign convention is used for both.

It is sometimes claimed that the characteristic waves are orthogonal \citep[e.g.,][]{k/m98}, which implies that their polarisation vectors would point in opposite directions on the Poincar\'e sphere. The radiative transfer is then approximated as a rotation of the polarisation vector of the emitted radiation around this axis \citep{k/m98,r/b02}. The polarisation of the characteristic waves are orthogonal when $\mathbf{E^1} \cdot \mathbf{E^{2^*}} = 0$ or $1+ K^1K^{2^*} = 0$. Hence, it is seen from Equation (\ref{eq8}) that a necessary condition for the characteristic waves to be orthogonal is $|K^{1,2}| = 1$. Likewise, from Equation (\ref{eq8})
\begin{equation}
	|K^{1,2}|^4 = 1 - \frac{4\rho_{\rm r}}{1+ |\rho|^2 + 2\rho_{\rm r}},
	\label{eq10}
\end{equation}
where the subscript "r" denotes the real part of $\rho$. In general then, the characteristic waves are not orthogonal and, hence, such a simplification should be used with care. However, one may note that when absorption is neglected, the characteristic waves will be orthogonal, since then $\rho_{\rm r} = 0$ (cf. Equation \ref{eq9}).

\section{Properties of the transfer equation}\label{sect3}

Although the transfer equation is trivial to solve when using characteristic waves (i.e., Equation \ref{eq6}), there are a few aspects of the solution that need to be emphasised. The polarisation properties of radiation are normally given in terms of the Stokes parameters and the emissivity ($\epsilon$) is specified for each one of them. Hence, the initial conditions in Equation (\ref{eq6}), i.e., $\mathbf{E}^{1,2}_{\rm o}$, need to be related to the emissivities of the individual Stokes parameters. This involves two steps: (1) Equation (\ref{eq6}) presupposes $100 \%$ polarised radiation. The emissivities should therefore be divided into two $100 \%$ polarised waves. (2) Each of these waves is then written as a sum of the two characteristic waves.

\subsection{Division into two characteristic waves}\label{sect3a}

Consider a $100 \%$ polarised wave, which initially has an electric field $\mathbf{E_{\rm o}}$ with polarisation $K_{\rm o} = E_{\rm y,o}/E_{\rm x,o}$. Its division into the two characteristic waves  $\mathbf{E}^{1,2}_{\rm o}$ yields 
\begin{eqnarray}
	E_{\rm x,o} & = & E^1_{\rm x,o} + E^2_{\rm x,o}\nonumber\\
	K_{\rm o}E_{\rm x,o} & = & K^1E^1_{\rm x,o} + K^2E^2_{\rm x,o},
	\label{eq11}
\end{eqnarray}
which can be solved to give
\begin{eqnarray}
	E^1_{\rm x,o} & = & -E_{\rm x,o}\frac{K_{\rm o}-K^2}{K^2 - K^1}\nonumber\\
	E^2_{\rm x,o} & = & E_{\rm x,o}\frac{K_{\rm o}-K^1}{K^2 - K^1}.
	\label{eq12}
\end{eqnarray}
The connection to the Stokes parameters is obtained from $|E_{\rm x,o}|^2 = (I_{\rm o} + Q_{\rm o})/2$ and $K^{*}_{\rm o} = (U_{\rm o} + iV_{\rm o})/2|E_{\rm x,o}|^2$. Without loss of generality $E_{\rm x,o}$ can be chosen to be real and one finds
\begin{eqnarray}
	E^1_{\rm x,o} & = & \sqrt{\frac{I_{\rm o}}{8(1+q_{\rm o})}}\left(1+q_{\rm o} - \frac{u_{\rm o}-
	iv_{\rm o}}		{K^2}\right)\nonumber\\
	E^2_{\rm x,o} & = & \sqrt{\frac{I_{\rm o}}{8(1+q_{\rm o})}}\left(1+q_{\rm o} + \frac{u_{\rm o}-
	iv_{\rm o}}		{K^2}\right),
	\label{eq13}
\end{eqnarray}
where $q_{\rm o}=Q_{\rm o}/I_{\rm o}$, $u_{\rm o}=U_{\rm o}/I_{\rm o}$, $v_{\rm o}=V_{\rm o}/I_{\rm o}$, and $K^1 = -K^2$ has been used.

As the wave propagates through the plasma its components vary according to $E_{\rm x} = E^1_{\rm x} + E^2_{\rm x}$ and $E_{\rm y} = K^1E^1_{\rm x} + K^2E^2_{\rm x} = K^2(-E^1_{\rm x} + E^2_{\rm x})$, where now $E^{1,2}_{\rm x} = E^{1,2}_{\rm x,o} \exp(-\kappa  s/2 \pm \Delta ks/2)$. With $U+iV = 2E_{\rm x}E^{*}_{\rm y}$, it is shown in Appendix A that after travelling a distance $s$, its circular polarisation is
\begin{eqnarray}
 \lefteqn{V =  I_{\rm o} \exp(-\kappa s)\left[ v_{\rm o} \left\{ \left(\frac{K_{\rm i}}{|K|}\right)^2 \cosh(\Delta k_{\rm r}s) +\left(\frac{K_{\rm r}}{|K|}\right)^2 \cos(\Delta k_{\rm i}s) \right\}\right.} \hspace{2.4cm}
	\nonumber\\
	& & - u_{\rm o}\frac{K_{\rm i}K_{\rm r}}{|K|^2}\left\{\cosh(\Delta k_{\rm r}s) - \cos(\Delta k_{\rm i}
	s)\right\} \nonumber\\
	& & + \frac{K_{\rm i}}{2}\left\{1 + \frac{1}{|K|^2} + q_{\rm o}\left(1- \frac{1}{|K|^2}\right)\right\}
	\sinh(\Delta k_{\rm r}s)	\nonumber\\
	& & + \left. \frac{K_{\rm r}}{2}\left\{1 - \frac{1}{|K|^2} + q_{\rm o}\left(1+ \frac{1}{|K|^2}\right)
	\right\}\sin(\Delta k_{\rm i}s) \right],
	\label{eq14}
\end{eqnarray} 	
where the subscripts "r" and "i" denote the real and imaginary parts, respectively, of a quantity. Furthermore, to simplify the notation, $K \equiv K^2$ has been introduced.

There are a number of general features of the circular polarisation that are apparent from Equation (\ref{eq14}), which will also be relevant for a homogeneous source, i.e., when emission occurs along the ray path. The things to note for a synchrotron source are: 

(1) The resulting value of $V$ depends linearly on the initial Stokes parameters ($v_{\rm o}$, $u_{\rm o}$, and $q_{\rm o}$). Since Stokes parameters are additive, although Equation (\ref{eq14}) was derived for a $100\%$ polarised wave, it is valid also in general for a partially polarised wave (i.e., $v_{\rm o}^2 + u_{\rm o}^2 +q_{\rm o}^2 < 1$). The same is true for the other Stokes parameters.

(2) The first term ($\propto v_{\rm o}$) corresponds to emission, while the second one ($\propto u_{\rm o}$) accounts for the conversion of linear to circular polarisation. This term is $\propto K_{\rm i}K_{\rm r}$, which implies a symmetric behaviour of $V$ for linear ($|K_{\rm i}| \ll 1$) and circular ($|K_{\rm r}| \ll 1$) characteristic waves (see Section \ref{sect4b} for further discussion of this issue). Although not obvious here, it will be shown later that the third and fourth terms correspond to absorption of the $U$ and $V$ parameters. Furthermore, the term $\ |K|^2 - 1$ explicitly shows the effects of the non-orthogonality of the characteristic waves. 

(3) It is seen from Equation (\ref{eq8}) that the magnitude of $\rho$ determines the polarisation properties of the characteristic waves; $|\rho| \ll 1$ corresponds to linearly polarised waves (i.e., $|K_{\rm i}| \ll 1$), while $|\rho| \gg 1$ corresponds to circularly polarised waves (i.e.,  $|K_{\rm r}| \ll 1$). 

(4) The magnitude of the transfer induced circular polarisation is largest when $|\rho| \sim 1$, since then $|K_{\rm r}| \sim| K_{\rm i}| \sim ||K|^2 -1| \sim 1$. An example of this can be seen in \cite{bjo90}, where the transition from linear to circular characteristic waves was discussed; in particular, it was shown that the degree of circular polarisation can reach several tens of percent, i.e., of the same order as the linear polarisation. It should also be noticed that the overall magnitude of the circular polarisation is determined by the polarisation properties of the characteristic waves (i.e., $K$), while its variation with frequency/optical depth is mainly due to their phase difference (i.e., $\Delta k$).

(5) It is seen that the sign chosen for $\sqrt{\Upsilon_{\rm V}^2 + \Upsilon_{\rm L}^2}$ is unimportant as long as it applies to both $K^{1,2}$ and $\Delta k$ (cf. Equation \ref{eq8}). 

\section{Circular polarisation from a homogeneous source}\label{sect4} 

The circular polarisation from a homogeneous source is obtained by integrating Equation (\ref{eq14}) from $s =  0$ to $s = s_{\rm max}$. This is done in Appendix B. Here, $s_{\rm max}$ is the thickness of the source so that its optical depth is $\tau = \kappa s_{\rm max}$. The observed circular polarisation in compact radio sources is usually of the order of one percent or smaller. Although inhomogeneities along a given sightline can severely affect the polarisation (this will be discussed in a forthcoming paper), the simplest explanation is that the physical conditions are such that either $|\rho| \ll 1$ or $|\rho| \gg 1$ (cf. the discussion in Section \ref{sect3a}). The plasma parameter with the least constrained value in compact radio sources is $\hat{\xi}_{\rm V}$. This is due to its sensitivity to two virtually unknown quantities, i.e., the number of low energy electrons (e.g., the low energy cut-off of the relativistic electrons) and the fraction of electron-positron pairs in the plasma (cf. Appendix C). Hence, the two limits of $|\rho|$ likely correspond to the two extremes $|\hat{\xi}_{\rm V}| \ll 1$ and $|\hat{\xi}_{\rm V}| \gg 1$.

\subsection{Circular polarisation from nearly circular characteristic waves}\label{sect4a}

It is convenient to write $\Delta k_{\rm r} s_{\rm max} \equiv \delta k_{\rm r} \tau$ and $\Delta k_{\rm i} s_{\rm max} \equiv \delta k_{\rm i} \tau$. When $|\Upsilon_{\rm V}| \gg |\Upsilon_{\rm L}|$, $|\rho| \gg 1$ and the characteristic waves are nearly circularly polarised (cf. Equation \ref{eq8}). In most cases this corresponds to $|\hat{\xi}_{\rm V}| \gg 1$. Expanding Equations (\ref{eq7}) and (\ref{eq8}) to lowest order in $|\rho|^{-1}$, one finds that $|K_{\rm r}|$, $|\delta k_{\rm r}|$, $|\delta k_{\rm i}|^{-1}$, and $|K|^2 - 1$ are all  $\sim |\rho|^{-1}$. Furthermore, let $v$ and $u$ denote the normalised $V$ and $U$ emissivities, respectively. Then $|v|$ and $|\xi_{\rm V}|$ are both small (cf. Appendix C). Assuming them to be of the same order of magnitude as  $|\rho|^{-1}$, the solution in Appendix B can be expanded to lowest order in  $\rho^{-1}$. This yields
\begin{eqnarray}
	V & = &S\left[v\left\{1-\exp(-\tau)\right\} - uK_{\rm r}\left\{1-\exp(-\tau)\right\}\right.\nonumber\\ 
	&+& \left. \delta k_{\rm r}\left\{1-\exp(-\tau)(1+ \tau)\right\}
	+  uK_{\rm r}\frac{\sin(\delta k_{\rm i}\tau)}{\delta k_{\rm i}}\exp(-\tau)\right],
	\label{eq15}
\end{eqnarray}
where $S  = \epsilon/\kappa$ is the source function. 

The relevant plasma parameters are $K_{\rm r} = -\hat{\xi}_{\rm U}/\hat{\xi}_{\rm V}$, $\delta k_{\rm r} = -(\xi_{\rm V} +\xi_{\rm U}\hat{\xi}_{\rm U}/\hat{\xi}_{\rm V}) = -\xi_{\rm V} + \xi_{\rm U} K_{\rm r}$, $\delta k_{\rm i} = \hat{\xi}_{\rm V}$, and $|K|^2 - 1 = -2\xi_{\rm U}/\hat{\xi}_{\rm V}$. With these expressions, it is straightforward to show that Equation (\ref{eq15}) is identical to the solution  given in \cite{bjo88}. However, to illuminate the various physical mechanisms influencing the observed circular polarisation a better representation of the solution is
\begin{eqnarray}
	V &=& S\left[(v - \xi_{\rm V})\left\{1-\exp(-\tau)(1+\tau)\right\} + v\tau \exp(-\tau)\right.\nonumber\\
	&-& \left. K_{\rm r}\left\{(u-\xi_{\rm U})\left\{1-\exp(-\tau)(1+\tau)\right\} + u\tau\exp(-\tau)\left(1 - 
	\frac{\sin(\hat{\xi}_{\rm V}\tau)}{\hat{\xi}_{\rm V}\tau}\right)\right\}\right].
	\label{eq16}
\end{eqnarray}
This shows explicitly the similarities between the $V$ and $U$ emissivities/absorptivities. For a thermal distribution of electrons, e.g., a relativistic Maxwellian, $v = \xi_{\rm V}$ and $u=\xi_{\rm U}$ \citep{j/h79}. Hence, it is the non-thermal aspect of the electron distribution which causes the change of sign in the circular polarisation at large optical depths. For a power law distribution of relativistic electrons both $|v - \xi_{\rm V}|$ and $|u-\xi_{\rm U}|$ are quite a bit smaller than $|v|$ and $|u|$, respectively \citep{j/o77}. Furthermore, for small optical depths, the non-thermal terms both vary as $\tau^2$, while the $v$- and $u$-terms vary as $\tau$ (for the $u$-term, this is valid for $\tau > \hat{\xi}_{\rm V}^{-1}$). Therefore, it is expected that for most electron distributions, the major contributions to the integrated circular polarisation come from the $v$- and $u$-terms. 

Although $q=0$ for synchrotron radiation, the $q$-term has been kept in the general solution given in Appendix B. The reason is to illustrate the nature of the conversion of linear to circular polarisation. It is sometimes said \citep{jon88, m/m18} that the conversion acts only on the Stokes parameter $Q$ and, hence, that the conversion in a synchrotron source occurs in two steps; first $U$ is converted to $Q$ through Faraday rotation and then $Q$ is converted to $V$. However, no $q$-term appears in Equation (\ref{eq16}). This implies that even if there were a $Q$-term, its contribution to the circular polarisation would be of order $\hat{\xi}_{\rm V}^{-2}$ and, hence, negligible. Another way of seeing the same thing is to consider the magnitude of the transfer induced circular polarisation, i.e., $|K_{\rm r}|$. Faraday rotation is $\propto \hat{\xi}_{\rm V}$ but $K_{\rm r} \propto \hat{\xi}_{\rm V}^{-1}$; i.e., larger Faraday rotation (larger $Q$) results in smaller circular polarisation. Hence, the name "Faraday conversion" often used for this process may be somewhat of a misnomer, since the conversion of $U$ to $V$ occurs directly without any intermediate steps. 

It is often assumed that absorption does not affect the conversion of $U$ to $V$ in the optically thin regime \citep[e.g.,][]{war98, ens03, osu13}. The solution to the transfer equation is then given by the Faraday conversion term, $V/I = u\tau_{\rm F}\tau_{\rm C}/6$, where $ \tau_{\rm F} = \hat{\xi}_{\rm V}\tau$ and $\tau_{\rm C} = \hat{\xi}_{\rm U}\tau$. However, expanding Equation (\ref{eq16}) for small optical depths and using $I=S\tau$, one finds for the conversion of linear to circular polarisation, $V/I = (\hat{\xi}_{\rm U}/\hat{\xi}_{\rm V})[\tau(u-\xi_{\rm U}) +u(1-\sin(\hat{\xi}_{\rm V}\tau)/\hat{\xi}_{\rm V}\tau)]$. A rapid rise in circular polarisation occurs at $\tau \sim \hat{\xi}_{\rm V}^{-1}$, so that for $\hat{\xi}_{\rm V} \tau > 1$, the leading term is $u\hat{\xi}_{\rm U}/\hat{\xi}_{\rm V}$. For $\hat{\xi}_{\rm V} \tau < 1$, the circular polarisation is substantially smaller, since it is determined by higher order terms. Among these is the Faraday conversion term, which is smaller by a factor $(\hat{\xi}_{\rm V} \tau)^2$ as compared to $u\hat{\xi}_{\rm U}/\hat{\xi}_{\rm V}$. Furthermore, the contribution from the non-thermal term ($\propto \tau (u-\xi_{\rm U})$)  may become significant, since it decreases with decreasing optical depth slower than the Faraday conversion term ($\tau$ vs $\tau^2$). It should also be noted that Equations (\ref{eq15}) and (\ref{eq16}) are correct only to first order in $|\rho|^{-1} \sim |\hat{\xi}_{\rm U}/\hat{\xi}_{\rm V}|$. Hence, second order terms may dominate the Faraday conversion term ($|\hat{\xi}_{\rm U}/\hat{\xi}_{\rm V}|$ vs $(\hat{\xi}_{\rm V} \tau)^2$). In this frequency range, \cite{j/o77} assumed that  the observed circular polarisation is unaffected by transfer effects and, instead, given directly by the emission process. Therefore, it is unlikely that neglect of absorption is a viable approximation even in the optically thin regime (cf., the discussion of the effects of absorption on the non-orthogonality of the characteristic waves at the end Section \ref{sect2a}). In general then, the Faraday conversion term does not provide a good approximation to the transfer induced circular polarisation. This aspect of the circular polarisation is discussed further in Section \ref{sect4b}.

\subsection{Circular polarisation from nearly linear characteristic waves}\label{sect4b}

When $|\Upsilon_{\rm V}| \ll |\Upsilon_{\rm L}|$, the characteristic waves are nearly linearly polarised, since $|\rho| \ll 1$. In this limit, $|K_{\rm i}|$ and $|K|^2-1$ are both $\sim |\rho|$. Since $|\xi_{\rm U}| \sim 1$ and $|\hat{\xi}_{\rm U}|$ cannot be assumed to be large in general, this corresponds in most cases to $|\hat{\xi}_{\rm V}| \ll 1$. The rather simple form of Equations (\ref{eq15}) and (\ref{eq16}) is due mainly to the properties of the phase difference $\delta k$ (i.e., $|\delta k_{\rm r}| \ll 1$ and $|\delta k_{\rm i}| \gg 1$). Here, on the other hand, one finds to lowest order in $\rho$, $\delta k_{\rm r} = -\xi_{\rm U}$ and $\delta k_{\rm i} = \hat{\xi}_{\rm U}$, the magnitude of which are both expected to be of order unity. This leads to a somewhat more complex expression for $V$. Expanding the solution in Appendix B to lowest order in $\rho$ yields 
\begin{eqnarray}
	V &=& S\left[\frac{(v - \xi_{\rm V} - q\hat{\xi}_{\rm U})}{1+\hat{\xi}_{\rm U}^2}\left\{1-\exp(-\tau)\left(\cos(\hat{\xi}_{\rm U}\tau) + \frac{\sin(\hat{\xi}_{\rm U}\tau)}{\hat{\xi}_{\rm U}}\right)\right\} + \frac{v\sin(\hat{\xi}_{\rm U}\tau) \exp(-\tau)}{\hat{\xi}_{\rm U}}\right.\nonumber\\
	&+&  K_{\rm i}\left\{\frac{(u-\xi_{\rm U})}{1-\xi_{\rm U}^2}\left(1-\exp(-\tau)\left(\cosh(\xi_{\rm U}\tau)+\frac{\sinh(\xi_{\rm U}\tau)}{\xi_{\rm U}}\right)\right) + \frac{u\sinh(\xi_{\rm U}\tau)\exp(-\tau)}{\xi_{\rm U}}\right\}\nonumber \\
	&-& \left.K_{\rm i}\left\{\frac{(u - \xi_{\rm U})}{1+\hat{\xi}_{\rm U}^2}\left(1-\exp(-\tau)\left(\cos(\hat{\xi}_{\rm U}\tau) + \frac{\sin(\hat{\xi}_{\rm U}\tau)}{\hat{\xi}_{\rm U}}\right)\right) + \frac{u\sin(\hat{\xi}_{\rm U}\tau) \exp(-\tau)}{\hat{\xi}_{\rm U}}\right\}\right], \nonumber\\
	\label{eq17}
\end{eqnarray}
where $K_{\rm i} = (\hat{\xi}_{\rm V}\hat{\xi}_{\rm U} + \xi_{\rm V}\xi_{\rm U})/(\hat{\xi}_{\rm U}^2 + \xi_{\rm U}^2)$ and $(|K|^2 - 1)/2 = (\xi_{\rm V}\hat{\xi}_{\rm U} - \hat{\xi}_{\rm V}\xi_{\rm U})/(\hat{\xi}_{\rm U}^2 + \xi_{\rm U}^2)$, which leads to $\hat{\xi}_{\rm U}(|K|^2 - 1)/2 = \xi_{\rm V} - \xi_{\rm U} K_{\rm i}$, have been used.

The structures of Equations (\ref{eq16}) and (\ref{eq17}) are rather similar. Although $|\delta k| \sim 1$ makes the variation of $V$ with $\tau$ more involved, their basic properties remain the same; e.g., the non-thermal terms (i.e., $v-\xi_{\rm V}$ and $u-\xi_{\rm U}$) are small compared to the $v$- and $u$-terms. They only become important at large optical depths, where they cause a change of sign. Furthermore, the amplitude of the conversion of linear to circular polarisation is determined by $K_{\rm i}$ rather than $K_{\rm r}$. 

This formal similarity between Equations (\ref{eq16}) and (\ref{eq17}) is due to the symmetric expressions of $\delta k$ and $K$ in the two limits (cf. Equations \ref{eq7} and \ref{eq8}). It is seen in Appendix B that the two limits can be related by just interchanging $\Upsilon_{\rm L}$ and $\Upsilon_{\rm V}$. Hence, several of the main properties, which derive from the birefringence of the plasma, can be obtained by interchanging $\hat{\xi}_{\rm U}$ and $\hat{\xi}_{\rm V}$. One example is the term giving the major contribution to the circular polarisation from conversion of linear polarisation. In Equation (\ref{eq16}) it is $\propto 1-\sin(\hat{\xi}_{\rm V}\tau)/\hat{\xi}_{\rm V}\tau$, while in Equation (\ref{eq17}) the corresponding expression is $\propto \sinh(\xi_{\rm U}\tau)/\xi_{\rm U}\tau - \sin(\hat{\xi}_{\rm U}\tau)/\hat{\xi}_{\rm U}\tau$. For nearly circular characteristic waves, this term causes a rapid rise in circular polarisation at $\tau \sim |\hat{\xi}_{\rm V}|^{-1}$ (cf. the discussion in Section \ref{sect4a}). Likewise, for nearly linear characteristic waves, this increase occurs at $\tau \sim |\hat{\xi}_{\rm U}|^{-1}$. The major difference is the values of $|\hat{\xi}_{\rm V}|$ and $|\hat{\xi}_{\rm U}|$ in the two limits. As already mentioned, they are expected to be quite different; while $|\hat{\xi}_{\rm V}| \gg 1$ for circular characteristic waves,  $|\hat{\xi}_{\rm U}|$ may not be much larger than unity for linear characteristic waves. Therefore, the observed circular polarisation is expected to come from, on average, larger optical depths for nearly linear as compared to nearly circular characteristic waves. 

The transition between these two limits was discussed in \cite{bjo90}. It was shown that the optical depth where the circular polarisation peaks decreases smoothly from $\tau >1$ to $\tau <1$ as the characteristic waves change from linear to circular \citep[see also][for the latter limit]{j/o77}. Physically, this can be understood as follows: When there are few low energy electrons, e.g., a power-law distribution of electrons with a low energy cut-off close to the synchrotron self-absorption frequency, $|\hat{\xi}_{\rm V}|\ll|\hat{\xi}_{\rm U}|$ and the characteristic waves are linearly polarised. As the low energy cut-off decreases, the magnitude of both $\hat{\xi}_{\rm V}$ and $\hat{\xi}_{\rm U}$ increase. The value of $\hat{\xi}_{\rm V}$ increases much faster than that of $\hat{\xi}_{\rm U}$, which causes the characteristic waves to change from linear to circular. At the same time, as the value of $\hat{\xi}_{\rm U}$ increases, so does the relative contribution to $V$ from the optically thin part of the spectrum (i.e., corresponding to $\tau \gsim |\hat{\xi}_{\rm U}|^{-1}$). 

This shows that the frequency distribution of the circular polarisation is expected to be quite different for nearly linear and nearly circular characteristic waves. Basically, this is due to the very different values of $\delta k_{\rm i}$ in the two cases; i.e., it is a consequence of the increase in circular polarisation at $\tau \sim |\delta k_{\rm i}|^{-1}$ together with an increasing value of $|\delta k_{\rm i}|$ as the characteristic waves change from nearly linear to nearly circular (cf. Equations \ref{eq7} and \ref{eq8}). The use of this property to distinguish observationally between linear and circular characteristic waves is discussed further in Section \ref{sect5}.

For nearly circular characteristic waves, the frequency range where $1> \tau> |\hat{\xi}_{\rm V}|^{-1}$ should be rather large. On the other hand, the corresponding frequency range for nearly linear characteristic waves is expected to be much smaller. Hence, $\tau< |\hat{\xi}_{\rm U}|^{-1}$ may dominate the optically thin region. Expanding Equation (\ref{eq17}) to lowest order in $\hat{\xi}_{\rm U}\tau$ gives
\begin{eqnarray}
	\frac{V}{I} &=& v(1-\tau) +(v-\xi_{\rm V})\frac{\tau}{2} + K_{\rm i}u(\xi_{\rm U}^2 +\hat{\xi}_{\rm U}^2)\frac{\tau^2}{6} - q\hat{\xi}_{\rm U}\frac{\tau}{2}\nonumber \\
	&=&  v(1-\tau) +(v-\xi_{\rm V})\frac{\tau}{2} +u\xi_{\rm V}\xi_{\rm U}\frac{\tau^2}{6} + u\hat{\xi}_{\rm V}\hat{\xi}_{\rm U}\frac{\tau^2}{6}  - q\hat{\xi}_{\rm U}\frac{\tau}{2}.
	\label{eq18} 
\end{eqnarray}

It is seen that the Faraday conversion term appears in Equation (\ref{eq18}), $u\hat{\xi}_{\rm V}\hat{\xi}_{\rm U}\tau^2/6 = u\tau_{\rm F}\tau_{\rm C}/6$. Since $\tau_{\rm F}\tau_{\rm C} = (\hat{\xi}_{\rm U}/\hat{\xi}_{\rm V})(\hat{\xi}_{\rm V}\tau)^2 = (\hat{\xi}_{\rm V}/\hat{\xi}_{\rm U})(\hat{\xi}_{\rm U}\tau)^2$, this appearance is another example of the symmetry between nearly circular and nearly linear characteristic waves. If this term were the dominant one, the name "Faraday conversion" would be appropriate in this limit. However, as discussed already in Section \ref{sect4a}, this is unlikely, since (1) it is really a third order term in the sense that both $\hat{\xi}_{\rm V}/\hat{\xi}_{\rm U}$ and $\hat{\xi}_{\rm U}\tau$ are much smaller than unity and (2) keeping also second order terms in the expansion parameter (i.e., $\hat{\xi}_{\rm V}/\hat{\xi}_{\rm U}$) could give contributions to $V$ larger than the Faraday conversion term.

In contrast to circular characteristic waves, Equation (\ref{eq17}) shows that a non-synchrotron $q$-term can affect the circular polarisation. Formally, the $q$-term is similar to the Faraday conversion term, since, roughly, $U\hat{\xi}_{\rm V}\tau/2$ is the $Q$-value produced by Faraday rotation of the synchrotron $U$-emission. Such an additional source of linearly polarised emission can give a significant contribution to $V$, since there is no restrictions on the value of $q$; for example, as shown by \cite{hod82}, this term can dominate the observed circular polarisation in inhomogeneous sources. 

\section{Discussion}\label{sect5}

The degree of circular polarisation observed in compact radio sources varies but it is rarely larger than $\sim 1\,\%$. This is roughly consistent with the level expected directly from the synchrotron emission process (see Appendix C). However, as discussed in the Introduction, there are reasons to believe that the observed circular polarisation is also affected by transport effects. This opens up a way to gain more detailed information about the source properties than is possible from the flux alone. As mentioned in Section \ref{sect4}, the rather low level of circular polarisation makes it likely that the characteristic waves are either nearly linearly or nearly circularly polarised. It is important to be able to distinguish between the two, since this has implications for some of the most central issues regarding the properties of compact radio sources; e.g., the presence of electron-positron pairs and the acceleration process of the radiating particles.

The flat spectrum of compact radio sources has been called a "cosmic conspiracy" by \cite{cot80}. \cite{b/k79} showed that a class of models in which relativistic electrons stream out in a jet with constant opening angle could account for the observations under two conditions: (1) The adiabatic losses of the electrons are compensated by a continuous re-acceleration so that their low energy cut-off stays constant. (2) The strength of the magnetic field varies inversely with radius. This leads to a constant brightness temperature along the jet. Such an inhomogeneous jet has a self-similar structure which implies no change of polarisation with frequency, since the parameters in the transfer equation stay constant along the jet.

Inhomogeneous sources can, roughly, be divided into two classes: (1) The source is homogeneous along each sightline but the source properties (e.g., the optical depth) vary between different sightlines. In its original form, the Blandford/K\"{o}nigl-jets belong to this class. (2) The source properties vary along a given sightline, e.g., due to turbulence. 

The first class of sources can be seen as a superposition of many homogeneous sources. In practice, the polarisation properties are obtained by integrating the results in Section \ref{sect4} over the appropriate range of parameter values. On the other hand, the polarisation properties of the second class of sources are more complicated to calculate, since, here, the characteristic waves don't propagate individually, i.e., they couple. In the present paper, it is assumed that the first kind of source model is sufficient to discuss the polarisation properties of compact radio sources. The effects of coupling of the characteristic waves will be treated in a forthcoming paper. 

Although the neglect of coupling could be seen as a serious limitation to the validity of the conclusions, this may not be so, at least not qualitatively. The reason is the following: It was mentioned in Section \ref{sect3a} that the amplitude of the circular polarisation is determined mainly by the polarisation properties of the characteristic waves ($K$), while its variation with frequency/optical depth is determined in large part by their phase difference ($\Delta k$). It is seen from Equation (\ref{eq3}) that the accumulated phase difference along a ray path does not depend on whether the medium is homogeneous or not. The coupling between the characteristic waves is due to variation of the local value of $K$ along the ray path. As a result, the coupling is expected to affect mainly the amplitude of the circular polarisation and less its frequency/optical depth dependence. This can be seen explicitly in \cite{bjo90}, where the circular polarisation from a homogeneous medium is compared to that emerging from a medium in which coupling is important. Therefore, the discussion below focuses on the frequency/optical depth dependence of the circular polarisation as a way to distinguish between nearly circular and nearly linear characteristic waves.

For flat spectrum radio sources, a substantial frequency dependence of the polarisation is expected only in the region around the spectral turnover, where the emission becomes optically thin. This occurs normally at rather large frequencies ($\sim 100\,$GHz) and it has only recently become possible to obtain high quality observations of the circular polarisation in this range for a fair number of sources \citep{thu17}. However, not all compact radio sources conform to the standard Blandford/K\"{o}nigl-jet model. Gigahertz-Peaked Spectrum sources are a class of objects, which have lower turnover frequencies ($\sim\,$few GHz) as well as a spectrum declining towards lower frequencies. It is clear that these sources are inhomogeneous, since, normally, their spectra are quite a bit flatter than the  characteristic $\nu^{5/2}$-spectrum of homogeneous sources. Hence, they are expected to show frequency dependent polarisation also in the optically thick part of the spectrum. A good example of such a source is PKS B2126-158 \citep{osu13}.

\subsection{The POLAMI survey}\label{sect5a}
In the POLAMI survey a large number of compact radio sources have been observed multiple times at 3 and 1.3\,mm \citep{agu17a}. The spectral index ($\alpha$) indicates that the flux is mostly optically thin radiation. However, there is a tendency for the spectrum to flatten when the flux increases \citep{agu17b}. This suggests that the turn-over frequency (i.e., $\tau \sim 1$) is, on average, close to 3\,mm. This sample is then a good starting point for a discussion of the origin of the circular polarisation. 

An important finding is that the maximum amplitude of circular polarisation is higher at 1\,mm (2.6\,\%) as compared to 3\,mm (2.0\,\%) \citep{thu17}. Furthermore, both of these values are, in turn, substantially larger than those found by others at longer wavelengths (i.e., optically thick frequencies). There are two implications from these observations which both suggest the presence of nearly circular characteristic waves. As shown in Section \ref{sect4}, the observed peak of the degree of circular polarisation in the optically thin regime is consistent with nearly circular characteristic waves but hard to reconcile with nearly linear characteristic waves. Also, in an inhomogeneous source, the polarisation at optically thick frequencies corresponds to an average over a range of optical depths. The sign change of the circular polarisation at large optical depths (due to the $u-\xi_{\rm U}$ term) is similar for both circular and linear characteristic waves. However, the relative contribution to the circular polarisation from this non-thermal term is larger for nearly circular characteristic waves as compared to the nearly linear ones, since $K_{\rm r} \propto \nu^{-1}$ and $K_{\rm i}\, \propsim \,\nu$ (cf. Equations \ref{eq16} and \ref{eq17}). Hence, the relative increase of the circular polarisation between the optically thick and thin parts of the spectrum should be larger for nearly circular characteristic waves as compared to nearly linear ones.

In the standard jet model, the spread in optical depth in the azimuthal direction is rather small and results in an averaging of possible rapid variations on small scales; cf. the integration over a thin shell done in \cite{j/o77}. Hence, observations of an unresolved source are determined mainly by the radial variations of the jet properties. 

For nearly circular characteristic waves, the circular birefringence is large (i.e., $|\hat{\xi}_{\rm V}|\gg 1$) and the main contribution to the linearly polarised flux comes from small optical depths, $\tau \sim |\hat{\xi}_{\rm V}|^{-1}$. Let $R_{\rm o}$ be the radius where $\tau = 1$ for some frequency $\nu$ and $\hat{\xi}_{\rm V, o}$ the corresponding value of $\hat{\xi}_{\rm V}$. With $B\propto R^{-1}$, the radial variation of the optical depth is $\tau = (R/R_{o})^{-(5+2\alpha)/2}$ and $\hat{\xi}_{\rm V}= \hat{\xi}_{\rm V,o}(R/R_{o})^{(1+2\alpha)/2}$. The radius where most of the linearly polarised flux is emitted ($R_{\rm L}$) is then obtained from $\tau |\hat{\xi}_{\rm V}|\sim 1$ as $R_{\rm L} \sim R_{\rm o} \hat{\xi}_{\rm V,o}^{1/2}$. Furthermore, the corresponding Stokes parameters are $U_{\rm L} \sim |Q_{\rm L}| \sim uS_{\rm o}  \hat{\xi}_{\rm V,o}^{-\alpha/2}$, where $S_{\rm o}$ is the source function at $R_{\rm o}$. Since, in this limit,  $U_{\rm L} \sim |Q_{\rm L}|$, small variations of the radial jet properties could give rise to rather large variations in the polarisation angle; in particular, this may account for the observed lack of a preferred polarisation angle in many sources \citep{agu17b}. 

If conversion from linear polarisation contributes significantly to the observed circular polarisation, one deduces $|\hat{\xi}_{\rm V,o}/\hat{\xi}_{\rm U,o}|\sim 10^2$ (cf. Equation \ref{eq16}). Assuming $|\hat{\xi}_{\rm U,o}| \sim 1$, $R_{\rm L} \sim 10 R_{\rm o}$ and the linearly polarised flux comes from a radius much larger than that of either the total or the circularly polarised flux. In line with observations, this implies that variations in linear polarisation should have a longer time scale than and correlated weakly with those in circular polarisation or total flux. Furthermore, with $\alpha \approx 1$, the depolarisation would also be $\sim10$, which shows that Faraday rotation could be responsible for a larger fraction of the observed depolarisation of the linear flux. 

The total and circularly polarised fluxes come, roughly, from the same region (i.e., $\tau \sim 1$). However, their sensitivity to changes in the various plasma parameters are very different. Not only does the circular polarisation vary more rapidly with optical depth than the total flux but, most importantly, the circular polarisation is sensitive to variations in plasma parameters that leave the total flux unaffected. As an example, for nearly circular characteristic waves, the magnitude of the circular polarisation due to conversion from linear polarisation is $\sim\hat{\xi}_{\rm U}/\hat{\xi}_{\rm V} \propto \gamma_{\rm i}^3/\ln \gamma_{\rm i}$ (Appendix C), where $ \gamma_{\rm i}$ is the Lorentz factor at the lower cut-off in the energy distribution of the relativistic electrons. The more rapid and uncorrelated variations of the circular flux as compared to the total flux observed by POLAMI could then come from small changes in $\gamma_{\rm i}$.

\subsection{PKS B2126-158} \label{5b}
\cite{osu13} have presented high quality, multi-frequency polarisation measurements of PKS B2126-158, which has a  turn-over frequency at 5.7 GHz. The source is inhomogeneous, since it has an inverted spectrum below this frequency ($\propsim\, \nu$). This makes it an ideal object for frequency dependent polarisation studies; in particular, in contrast to the flat spectrum sources, frequency dependent polarisation is expected in the optically thick part of the spectrum. The circular polarisation peaks at a frequency above the turn-over frequency, indicating nearly circular characteristic waves. Several of its properties are as expected for a homogeneous source with $|\hat{\xi}_{\rm V}|\gg 1$ \citep[see][]{j/o77}; for example, a broad minimum in the degree of linear polarisation coincide with the maximum in circular polarisation and there is a clear indication of a $\sim 90^{o}$ swing in the polarisation angle in the optically thick part of the spectrum (i.e., $Q$ changes sign).

However, there are two aspects of the observations which do not fit with a homogeneous source, namely, the lack of a sign change of the circular polarisation in the optically thick part of the spectrum and the apparently smooth $\sim 90^{o}$ swing in the polarisation angle rather than an abrupt flip. In order to see how these can be accounted for by an inhomogeneous source structure, a few of its properties needs to be considered. 

The range of optical depths in an inhomogeneous source, which contributes to the flux at a given frequency, depends on the slope in the optically thick part of the spectrum. For a flat spectrum, the polarisation is independent of frequency and no change of sign is observed in either $Q$ or $V$. As the spectrum becomes more inverted, the relative importance of the large optical depths increases. Hence, for some value of the slope, sign changes will be observed for $Q$ and/or $V$.
For $\tau |\hat{\xi}_{\rm V}| > 1$, it can be shown that in a homogeneous source $Q = S[\xi_{\rm U}(1-\exp(-\tau)) - u]/\hat{\xi}_{\rm V}$. As compared to the circular polarisation in Equation (\ref{eq16}), there are two differences: (1) The sign change in $Q$ occurs at smaller optical depth than the corresponding change for $V$. (2) Since $Q \propto \hat{\xi}_{\rm V}^{-1}$ and  $V \propto \hat{\xi}_{\rm U}/\hat{\xi}_{\rm V}$, the amplitude of $Q$ decreases with frequency somewhat faster than does the one for $V$ ($\nu^{-1.2}$ vs $\nu^{-1}$, where $\alpha = 0.7$ has been used). Hence, the relative importance of large optical depths is larger for $Q$ than for $V$. Both of these effects cause the sign change in $Q$ to occur at a higher frequency than for $V$. In a forthcoming paper, it will be discussed how the observed change of sign in $Q$ but not in $V$ can be made consistent with the observed spectrum.

In general then, sign changes in $V$ and $Q$ in inhomogeneous sources depend on the slope in the optically thick part of the  spectrum. Observations with high spatial resolution may resolve some of the inhomogeneities and, hence, make it more likely to find such sign changes. This could be the case for the VLBA-observations of NGC\,1275 (3C\,84), where the sign of the circular polarisation changed between the optically thick and thin parts of the source \citep{h/w04}. Unfortunately, no linear polarisation was detected so the expected concurrent sign change in $Q$ could not be established.

In contrast to the circular polarisation, the value of $Q$ is determined by contributions from two very different regions in the jet. The optically thin emission comes from radii much larger than that at $\tau \sim 1$, which emits $Q$-flux with opposite sign. As the spectrum becomes increasingly inverted, the relative contribution to $Q$ from the optically thin emission goes down. As mentioned above, $U \sim |Q|$ for this component even when the total $Q$ changes sign and, hence, the value of $U$ will be non-negligible. This causes the $90^{o}$ flip in position angle observed in a homogeneous source to be replaced by a smooth $90^{o}$ swing in an inhomogeneous source.

\subsection{Observational implications} \label{sect5c}
Both the POLAMI sample and the detailed observations of PKS B2126-158 are most straightforwardly understood for characteristic waves, which are nearly circular polarised. This conclusion rests on the observed frequency distribution of the circular polarisation and implies $|\hat{\xi}_{\rm V}/\hat{\xi}_{\rm U}| \gg 1$. Its actual value is harder to estimate, since, as discussed above, the magnitude of the circular polarisation may be seriously affected by inhomogeneities along various lines of sight. However, the properties of the linear polarisation in the POLAMI sample can be accounted for by a value of $|\hat{\xi}_{\rm V}|$ consistent with only minor contributions from inhomogeneities. Assuming this to be the case, the value of $|\hat{\xi}_{\rm V}/\hat{\xi}_{\rm U}|\sim 10^2$ can be used to constrain the properties of the synchrotron plasma.

In addition to the magnetic field direction, when the frequency dependence of the transfer coefficients are normalised to the turn-over frequency, there are two free parameters (see Appendix C); namely, $\gamma_{\rm min}$ and the number of electron-positron pairs ($n_{\rm p}$) relative the excess number of electrons ($n_{\rm exc}$). With $|\hat{\xi}_{\rm V}/\hat{\xi}_{\rm U}|\sim 10^2$, one finds $(\gamma_{\rm min}^3/\ln \gamma_{\rm min})(1+2n_{\rm p}/n_{\rm exc}) \sim 10^2$ (Equation \ref{c4}). Although the presence of nearly circular characteristic waves by itself is enough to show that $\gamma_{\rm min}$ is much below that corresponding to the turn-over frequency (i.e., $\gamma_{\rm min} \ll \gamma_{\rm abs} \approx 10^2$, see Appendix C), observations allow a fair fraction of electron-positron pairs. An upper limit from the relativistic particles is obtained for $\gamma_{\rm min} \sim 1$, i.e., the particles are injected into the acceleration process with trans-relativistic energies. This gives $n_{\rm p}/n_{\rm exc}\,\lsim\,10^2$. The emission coefficient for the circular polarisation depends on $n_{\rm exc}/n_{\rm p}$ but not  $\gamma_{\rm min}$. Hence, the degeneracy between the two can be broken by direct observation of the circular polarisation intrinsic to the synchrotron process. However, this may require observations in the frequency range corresponding to $|\hat{\xi}_{\rm V}|\tau\,\lsim\,1$ (see also below).

The conversion of linear to circular polarisation is often described by the Faraday conversion term $u\tau_{\rm F}\tau_{\rm C}/6$, which has a very steep frequency dependence ($\propto \nu^{-5}$). Since observations indicate a more modest frequency dependence of the circular polarisation, this has limited more detailed modelling of the sources properties \citep[e.g.,][]{osu13,thu17}. However, it was shown in Section \ref{sect4} that this term is unlikely to significantly affect the observed circular polarisation. Instead, as argued above, the use of the full solution to the transfer equation allows a rather direct interpretation of the observations.

Nearly circular characteristic waves imply large Faraday depths over a wide range of frequencies. The apparent lack of observed Faraday rotation has been used to argue, instead, that the characteristic waves are linearly polarised \citep{war77}. Although, in the standard jet model, polarisation in the flat, optically thick  part of the spectrum should be constant, Faraday rotation is expected in the optically thin part. However, even here, the polarisation angle should remain constant until the transition to the Faraday thin regime occurs (i.e., $|\hat{\xi}_{\rm V}|\tau \sim 1$). With the source parameters deduced above from the observed circular polarisation (e.g., $|\hat{\xi}_{\rm V}| \sim 10^2$), this transition takes place at a frequency $|\hat{\xi}_{\rm V}|^{1/2} \sim 10$ larger than the turn-over frequency. Accurate polarisation measurements may be hard to obtain at such frequencies. Furthermore, the change in position angle should be smaller than for a homogeneous source. Since $U \sim |Q|$, the change in position angle is expected to be $\sim \pi/8$ rather than $\sim \pi/4$ for a homogeneous source.

\section{Conclusions}\label{sect6}

The transfer equation of polarised light in a homogeneous medium can be solved analytically. However, the standard solution is complex and observations are usually discussed in terms of various approximations. The main conclusions in the present paper are:

1) The use of characteristic waves allows an alternative way of expressing the transfer equation. The solution is more compact and transparent regarding the physical mechanisms determining the emerging polarisation than in the standard formulation.

2) The frequency dependence of the circular polarisation is a direct way of establishing the properties of the characteristic waves.

3) High quality observations of circular polarisation in compact radio sources indicate that the characteristic waves are nearly circularly polarised. This provides, for example, an upper limit to the fraction of electron-positron pairs.

4) Several of the approximations in common use have limited applicability; for example, it is shown that the Faraday conversion term is unlikely to have a significant impact on the observed circular polarisation.

\newpage

\appendix

\begin{center}
{\bf Appendix}
\end{center}

\section{Propagation of a polarised light ray}
With the initial conditions given by Equation (\ref{eq13})
\begin{eqnarray}
U+iV &=& 2E_{\rm x}E_{\rm y}^{*} = \frac{I_{\rm o}K^{*} \exp(-\kappa s)}{4(1+q_{\rm o})}\nonumber \\
	&\times& \left[\left\{(1+q_{\rm o} - \sigma)\exp(\frac{\Delta k}{2}s) + (1+q_{\rm o} +\sigma)\exp(\frac{-\Delta k}{2}s)\right\}\right. \nonumber \\
	&\times& \left.\left\{-(1+q_{\rm o} -\sigma^{*})\exp(\frac{\Delta k^{*}}{2}s)+ (1+q_{\rm o} +\sigma^{*})\exp(\frac{-\Delta k^{*}}{2}s)\right\}\right],
	\label{a1}
\end{eqnarray}
where, again, $K^2 \equiv K$ has been used together with $\sigma \equiv (u_{\rm o} - iv_{\rm o})/K$. The terms in Equation (\ref{a1}) can be rearranged to give
\begin{eqnarray}
U+iV = \left.\frac{I_{\rm o}K^{*} \exp(-\kappa s)}{4(1+q_{\rm o})}\right[
	&-&\exp(\Delta k_{\rm r}s) \left\{(1+q_{\rm o})^2 + |\sigma|^2 - (1+q_{\rm o}) (\sigma^{*} + \sigma)\right\}\nonumber \\
	  &+&\exp(-\Delta k_{\rm r}s) \left\{(1+q_{\rm o})^2 + |\sigma|^2 + (1+q_{\rm o}) (\sigma^{*} + \sigma^{*})\right\}\nonumber \\
	  &+&\exp(i\Delta k_{\rm i}s) \left\{(1+q_{\rm o})^2 - |\sigma|^2 + (1+q_{\rm o}) (\sigma^{*} - \sigma)\right\}\nonumber \\
	  &-&\left.\exp(-i\Delta k_{\rm i}s) \left\{(1+q_{\rm o})^2 - |\sigma|^2 - (1+q_{\rm o}) (\sigma^{*} - \sigma)\right\}\right]\nonumber \\.
	  \label{a2}
\end{eqnarray}
 Since the wave is $100\%$ polarised, $|\sigma|^2 = (1-q_{\rm o}^2)/|K|^2$ and Equation (\ref{a2}) can be rewritten as
\begin{eqnarray}
\lefteqn{U+iV =I_{\rm o}K^{*} \exp(-\kappa s)
		\left[\sigma_{\rm r}\cosh(\Delta k_{\rm r}s) - \left(\frac{|K|^2+1}{2|K|^2} + 
		q_{\rm o}\frac{|K|^2-1}{2|K|^2}\right) \sinh(\Delta k_{\rm r}s)\right.} 
		\hspace{4cm}
	        \nonumber\\
		& & -i \left.\left\{\sigma_{\rm i}\cos(\Delta k_{\rm i}s) - \left(\frac{|K|^2-1}{2|K|^2} +
		q_{\rm o}\frac{|K|^2+1}{2|K|^2}\right) \sin(\Delta k_{\rm i}s) \right\}\right]. \nonumber \\
		\label{a3}
\end{eqnarray}
With the use of
\begin{eqnarray}
	\sigma &=& \frac{K^{*}(u_{\rm o} - iv_{\rm o})}{|K|^2} \nonumber \\
	            &=& \frac{(K_{\rm r}u_{\rm o} - K_{\rm i}v_{\rm o}) -i(K_{\rm i}u_{\rm o} + K_{\rm r}v_{\rm o})}{|K|^2},
	  \label{a4}
\end{eqnarray}
one finds 
\begin{eqnarray}
\lefteqn{V = I_{\rm o}\exp(-\kappa s) \left[ v_{\rm o}\left\{\left(\frac{K_{\rm i}}{|K|}\right)^2 \cosh(\Delta 		k_{\rm r}s) + \left(\frac{K_{\rm r}}{|K|}\right)^2 \cos(\Delta k_{\rm i}s)\right\}\right.}
	\hspace{2.5cm}
	\nonumber \\
	& &+\frac{u_{\rm o}K_{\rm i}K_{\rm r}}{|K|^2} \left\{-\cosh(\Delta k_{\rm r}s) +\cos(\Delta k_{\rm i}	s)\right\} \nonumber \\
	& &+\frac{K_{\rm i}}{2}\left\{\frac{|K|^2 + 1}{|K|^2 } +q_{\rm o}\left(\frac{|K|^2 - 1}{|K|^2 }\right)		\right\}\sinh(\Delta k_{\rm r}s) \nonumber \\
	& &+\left.\frac{K_{\rm r}}{2}\left\{\frac{|K|^2 - 1}{|K|^2 } +q_{\rm o}\left(\frac{|K|^2 + 1}{|K|^2 }		\right)\right\}\sin(\Delta k_{\rm i}s)\right],
	\label{a5}
\end{eqnarray}
and
\begin{eqnarray}
\lefteqn{U = I_{\rm o}\exp(-\kappa s) \left[ u_{\rm o}\left\{\left(\frac{K_{\rm r}}{|K|}\right)^2 \cosh(\Delta k_{\rm r}s) + \left(\frac{K_{\rm i}}{|K|}\right)^2 \cos(\Delta k_{\rm i}s)\right\}\right.}
	\hspace{2.5cm}
	\nonumber \\
	& &+\frac{v_{\rm o}K_{\rm i}K_{\rm r}}{|K|^2} \left\{-\cosh(\Delta k_{\rm r}s) +\cos(\Delta k_{\rm i}	s)\right\} \nonumber \\
	& &-\frac{K_{\rm r}}{2}\left\{\frac{|K|^2 + 1}{|K|^2 } +q_{\rm o}\left(\frac{|K|^2 - 1}{|K|^2 }\right)		\right\}\sinh(\Delta k_{\rm r}s) \nonumber \\
	& &+\left.\frac{K_{\rm i}}{2}\left\{\frac{|K|^2 - 1}{|K|^2 } +q_{\rm o}\left(\frac{|K|^2 + 1}{|K|^2 }		\right)\right\}\sin(\Delta k_{\rm i}s)\right].
	\label{a6}
\end{eqnarray}

Likewise
\begin{eqnarray} 
|E_{\rm x}|^2 &=&  \frac{I_{\rm o}\exp(-\kappa s)}{8(1+q_{\rm o})}\nonumber \\
	&\times& \left[\left\{(1+q_{\rm o} - \sigma)\exp(\frac{\Delta k}{2}s) + (1+q_{\rm o} +\sigma)\exp(\frac{-\Delta k}{2}s)\right\}\right. \nonumber \\
	&\times& \left.\left\{(1+q_{\rm o} -\sigma^{*})\exp(\frac{\Delta k^{*}}{2}s)+ (1+q_{\rm o} +\sigma^{*})\exp(\frac{-\Delta k^{*}}{2}s)\right\}\right],
	\label{a7}
\end{eqnarray}	
and
\begin{eqnarray} 
	|E_{\rm y}|^2 &=&  \frac{I_{\rm o}|K|^2\exp(-\kappa s)}{8(1+q_{\rm o})}\nonumber \\
	&\times& \left[\left\{-(1+q_{\rm o} - \sigma)\exp(\frac{\Delta k}{2}s) + (1+q_{\rm o} +\sigma)			\exp(\frac{-\Delta k}{2}s)\right\}\right. \nonumber \\
	&\times& \left.\left\{-(1+q_{\rm o} -\sigma^{*})\exp(\frac{\Delta k^{*}}{2}s)+ (1+q_{\rm o} +			\sigma^{*})\exp(\frac{-\Delta k^{*}}{2}s)\right\}\right],
	\label{a8}
\end{eqnarray}
which leads to
\begin{eqnarray}
\lefteqn{|E_{\rm x}|^2 =\frac{I_{\rm o}\exp(-\kappa s)}{2}
		\left[\left(\frac{|K|^2+1}{2|K|^2} + 
		q_{\rm o}\frac{|K|^2-1}{2|K|^2}\right)\cosh(\Delta k_{\rm r}s) -  \sigma_{\rm r}\sinh(\Delta 			k_{\rm r}s)\right.} 
		\hspace{5.2cm}
	        \nonumber\\
		& &\hspace{-1cm} +\left.\left(\frac{|K|^2-1}{2|K|^2} +
		q_{\rm o}\frac{|K|^2+1}{2|K|^2}\right)\cos(\Delta k_{\rm i}s) +  \sigma_{\rm i}\sin(\Delta 			k_{\rm i}s)\right]. \nonumber \\
		\label{a9}
\end{eqnarray}	
and
\begin{eqnarray}
\lefteqn{|E_{\rm y}|^2 =\frac{I_{\rm o}|K|^2\exp(-\kappa s)}{2}
		\left[\left(\frac{|K|^2+1}{2|K|^2} + 
		q_{\rm o}\frac{|K|^2-1}{2|K|^2}\right)\cosh(\Delta k_{\rm r}s) -  \sigma_{\rm r}\sinh(\Delta 			k_{\rm r}s)\right.} 
		\hspace{4.5cm}
	        \nonumber\\
		& &\hspace{-0.5cm} -\left.\left(\frac{|K|^2-1}{2|K|^2} +
		q_{\rm o}\frac{|K|^2+1}{2|K|^2}\right)\cos(\Delta k_{\rm i}s) -  \sigma_{\rm i}\sin(\Delta        			k_{\rm i}s)\right].\nonumber \\
		\label{a10}
\end{eqnarray}
With the use of Equation (\ref{a4}), Equations (\ref{a9}) and (\ref{a10}) can be combined to give	            
\begin{eqnarray}
Q \equiv |E_{\rm x}|^2 - |E_{\rm y}|^2 &=& I_{\rm o} \exp(-\kappa s)\nonumber \\
	&\times&\left[ u_{\rm o}\left\{\frac{K_{\rm r}(|K|^2-1)}{2|K|^2} \sinh(\Delta k_{\rm r}s) - 				\frac{K_{\rm i}(|K|^2+1)}{2|K|^2} \sin(\Delta k_{\rm i}s)\right\}\right.
	\nonumber \\
	&&-v_{\rm o}\left\{\frac{K_{\rm i}(|K|^2-1)}{2|K|^2} \sinh(\Delta k_{\rm r}s) + \frac{K_{\rm r}(|K|		^2+1)}{2|K|^2} \sin(\Delta k_{\rm i}s)\right\}\nonumber \\
	& &-\left\{\frac{|K|^4 - 1}{4|K|^2} +q_{\rm o}\frac{(|K|^2 - 1)^2}{4|K|^2}\right\}\cosh(\Delta k_{\rm 		r}s) \nonumber \\
	& &+\left.\left\{\frac{|K|^4 - 1}{4|K|^2} +q_{\rm o}\frac{(|K|^2 + 1)^2}{4|K|^2}\right\}\cos(\Delta 		k_{\rm i}s)\right]
	\label{a11}
\end{eqnarray}
and
\begin{eqnarray}
I \equiv |E_{\rm x}|^2 + |E_{\rm y}|^2 &=& I_{\rm o} \exp(-\kappa s)\nonumber \\
	&\times&\left[ u_{\rm o}\left\{-\frac{K_{\rm r}(|K|^2+1)}{2} \sinh(\Delta k_{\rm r}s) + \frac{K_{\rm i}(|		K|^2-1)}{2} \sin(\Delta k_{\rm i}s)\right\}\right.
	\nonumber \\
	&&+v_{\rm o}\left\{\frac{K_{\rm i}(|K|^2+1)}{2} \sinh(\Delta k_{\rm r}s) + \frac{K_{\rm r}(|K|^2-1)}		{2} \sin(\Delta k_{\rm i}s)\right\}\nonumber \\
	& &+\left\{\frac{(|K|^2 + 1)^2}{4|K|^2} +q_{\rm o}\frac{|K|^4 - 1}{4|K|^2}\right\}\cosh(\Delta k_{\rm 	r}s) \nonumber \\
	& &-\left.\left\{\frac{(|K|^2 - 1)^2}{4|K|^2} +q_{\rm o}\frac{|K|^4 -1}{4|K|^2}\right\}\cos(\Delta 		k_{\rm i}s)\right],
	\label{a12}
\end{eqnarray}
where the terms have been grouped so as to emphasise the various physical mechanisms at play.		            	

\section{General solution to the transfer equation}
For a homogeneous source, Equations (\ref{a5}), (\ref{a6}), (\ref{a11}), and (\ref{a12}) need to be integrated through the emission region from $s=0$ to $s= s_{\rm max}$, where $s_{\rm max}$ is the thickness of the source. Furthermore, the intensity is replaced by the emissivity ($\epsilon$) so that $I_{\rm o} \rightarrow \epsilon \,{\rm d}s = S\,{\rm d}\tau$, where $S \equiv \epsilon/\kappa$ is the source function. The polarisation for a homogeneous source is then obtained directly from Equations (\ref{a5}), (\ref{a6}), (\ref{a11}), and (\ref{a12}) by substituting
\begin{eqnarray}
	\exp(-\kappa s)\cosh(\Delta k_{\rm r}s) &\rightarrow& \frac{1 - \exp(-\tau)\{\delta k_{\rm r} \sinh(\delta k_{\rm r}\tau) + \cosh(\delta k_{\rm r}\tau)\}}{1-\delta k_{\rm r}^2} \nonumber \\
	\exp(-\kappa s)\sinh(\Delta k_{\rm r}s) &\rightarrow& \frac{\delta k_{\rm r} - \exp(-\tau)\{\delta k_{\rm r} \cosh(\delta k_{\rm r}\tau) + \sinh(\delta k_{\rm r}\tau)\}}{1-\delta k_{\rm r}^2} \nonumber \\
	\exp(-\kappa s)\cos(\Delta k_{\rm i}s) &\rightarrow& \frac{1 - \exp(-\tau)\{-\delta k_{\rm i} \sin(\delta k_{\rm i}\tau)+ \cos(\delta k_{\rm i}\tau)\}}{1+\delta k_{\rm i}^2} \nonumber \\
	\exp(-\kappa s)\sin(\Delta k_{\rm i}s) &\rightarrow& \frac{\delta k_{\rm i} - \exp(-\tau)\{\delta k_{\rm i} \cos(\delta k_{\rm i}\tau)+ \sin(\delta k_{\rm i}\tau)\}}{1+\delta k_{\rm i}^2}, \nonumber \\
	\label{b1}
\end{eqnarray}
where $\tau = \kappa s_{\rm max}$ is the optical depth of the source, $\delta k_{\rm r} = \Delta k_{\rm r}/\kappa$, and $\delta k_{\rm i} = \Delta k_{\rm i}/\kappa$.

The result for the circular polarisation is
\begin{eqnarray}
	\lefteqn{V=\frac{S}{|K|^2}\left[\frac{K_{\rm i}}{1-\delta k_{\rm r}^2}\left\{-uK_{\rm r} + vK_{\rm i} + \delta k_{\rm r}\left(\frac{|K|^2 +1}{2} + q\frac{|K|^2 -1}{2}\right)\right\}\right.}
	\hspace{2cm} \nonumber \\
	& & \hspace{-1cm}+\frac{K_{\rm r}}{1+\delta k_{\rm i}^2}\left\{uK_{\rm i} + vK_{\rm r} + \delta k_{\rm i}\left(\frac{|K|^2 -1}{2} + q\frac{|K|^2 +1}{2}\right)\right\}
	\nonumber \\
	\lefteqn{-\exp(-\tau)\left\{\frac{K_{\rm i}}{1-\delta k_{\rm r}^2}\left(-uK_{\rm r} + vK_{\rm i} + \delta k_{\rm r}\left(\frac{|K|^2 +1}{2} + q\frac{|K|^2 -1}{2}\right)\right)\cosh(\delta k_{\rm r}\tau)\right.} \hspace{1cm}\nonumber \\
	& &\hspace{0.25cm}+\frac{K_{\rm i}}{1-\delta k_{\rm r}^2}\left(\delta k_{\rm r}(-uK_{\rm r} +vK_{\rm i}) + \frac{|K|^2 +1}{2} + q\frac{|K|^2 -1}{2}\right)\sinh(\delta k_{\rm r}\tau) 
	\nonumber \\
	& &\hspace{0.25cm}+\frac{K_{\rm r}}{1+\delta k_{\rm i}^2}\left(uK_{\rm i} + vK_{\rm r} + \delta k_{\rm i}\left(\frac{|K|^2 -1}{2} + q\frac{|K|^2 +1}{2}\right)\right)\cos(\delta k_{\rm i}\tau) \nonumber \\
	& &\hspace{0.15cm}\left.\left.-\frac{K_{\rm r}}{1+\delta k_{\rm i}^2}\left(\delta k_{\rm i}(uK_{\rm i} +vK_{\rm r}) - \frac{|K|^2 -1}{2} - q\frac{|K|^2 +1}{2}\right)\sin(\delta k_{\rm i}\tau)\right\}\right], 
	\nonumber \\
	\label{b2}
\end{eqnarray}
where, now, the various terms have been grouped in order to emphasise the variation of the circular polarisation with optical depth.	

\subsection{Limiting solution for $|\rho|\gg 1$}
From Equation (\ref{eq7}), $\delta k = i\sqrt{\Upsilon_{\rm V}^2 + \Upsilon_{\rm L}^2} = i \Upsilon_{\rm V} \sqrt{1 + \Upsilon_{\rm L}^2/\Upsilon_{\rm V}^2}$. With $|\rho| \gg 1|$ and regarding $|\xi_{\rm V}|$ to be of the same order of smallness as $|\hat{\xi}_{\rm V}|^{-1}$, one finds to first order in $\Upsilon_{\rm L}/\Upsilon_{\rm V}$,  
\begin{eqnarray}
\delta k &=& i \left(\Upsilon_{\rm V} +  \frac{\Upsilon_{\rm L}^2}{2\Upsilon_{\rm V}}\right) \nonumber \\
	& = & -\left(\xi_{\rm V} +\frac{\xi_{\rm U}\hat{\xi}_{\rm U}}{\hat{\xi}_{\rm V}}\right) + i\hat{\xi}_{\rm V} .
	\label{b3}	
\end{eqnarray}			
Note that the real and imaginary parts are of different orders and that only the leading order term has been retained for each of them. The same practice is followed below. Likewise from Equation (\ref{eq8}), the polarisation of the characteristic wave $K^2 \equiv K = \delta k/(\Upsilon_{\rm V} -i\Upsilon_{\rm L})$ is 
\begin{eqnarray}
 	K &=& i\left(1 + i\frac{\Upsilon_{\rm L}}{\Upsilon_{\rm V}}\right) \nonumber \\
		&=&  -\frac{\hat{\xi}_{\rm U}}{\hat{\xi}_{\rm V}} + i\left(1 - \frac{\xi_{\rm U}}{\hat{\xi}_{\rm V}}\right). 
		\label{b4}
\end{eqnarray}
The degree of non-orthogonality between the characteristic waves is obtained from $(|K|^2 -1)/2 = -\xi_{\rm U}/\hat{\xi}_{\rm V}$.

It is seen from Equations (\ref{b3}) and (\ref{b4}) that $|\delta k_{\rm r}|$, $|\delta k_{\rm i}|^{-1}$, $|K_{\rm r}|$, and $(|K|^2 -1|$ are all of order $|\rho|^{-1}$. The circular polarisation to first order in  $|\rho|^{-1}$ is then obtained directly from Equation (\ref{b2}) as
\begin{equation}
	V= S\left[-uK_{\rm r} + v + \delta k_{\rm r} - \exp(-\tau)\left\{v + \delta k_{\rm r}(1+\tau)- uK_{\rm r}\left(1 + \frac{\sin(\delta k_{\rm i}\tau)}{\delta k_{\rm i}}\right)\right\}\right].
	\label{b5}
\end{equation}

\subsection{Limiting solution for $|\rho|\ll 1$}
When $|\rho| \ll 1$, one finds to first order in $\Upsilon_{\rm V}/\Upsilon_{\rm L}$,  
\begin{eqnarray}
\delta k &=& i \left(\Upsilon_{\rm L} +  \frac{\Upsilon_{\rm V}^2}{2\Upsilon_{\rm L}}\right) \nonumber \\
	& = & -\xi_{\rm U} + i \hat{\xi}_{\rm U}
	\label{b6}
\end{eqnarray}
No higher order terms are retained for $\delta k_{\rm r}$, since $ |\xi_{\rm U}| \sim 1$. The polarisation of the characteristic wave is given by
\begin{eqnarray}
	K &=& -1+\frac{i\Upsilon_{\rm V}}{\Upsilon_{\rm L}} \nonumber \\
	   &=& -1 + \frac{\hat{\xi}_{\rm V}\xi_{\rm U} - \xi_{\rm V}\hat{\xi}_{\rm U}}{\hat{\xi}_{\rm U}^2 + \xi_{\rm U}^2} + i \frac{\hat{\xi}_{\rm V}\hat{\xi}_{\rm U} + \xi_{\rm V}\xi_{\rm U}}{\hat{\xi}_{\rm U}^2 + \xi_{\rm U}^2}.
	   \label{b7}
\end{eqnarray}
Again, the degree of non-orthogonality of the characteristic waves is obtained from $(|K|^2 - 1)/2 = (\xi_{\rm V}\hat{\xi}_{\rm U} - \hat{\xi}_{\rm V}\xi_{\rm U})/(\hat{\xi}_{\rm U}^2 + \xi_{\rm U}^2)$. 

In this limit, the small quantities are $|K_{\rm i}|$ and $|K|^2 -1$, which are both $\sim |\rho|$. Expanding equation (\ref{b2}) to first order in $|\rho|$ yields
\begin{eqnarray}
	\lefteqn{V=S\left[\frac{K_{\rm i}}{1-\delta k_{\rm r}^2} (u + \delta k_{\rm r}) +\frac{1}{1+\delta k_{\rm i}^2}\left(-uK_{\rm i} +v - \delta k_{\rm i}\left(\frac{|K|^2-1}{2} + q\right)\right)\right.}
	 \hspace{12cm}\nonumber \\
	 \lefteqn{-\exp(-\tau)\left\{\frac{K_{\rm i}}{1-\delta k_{\rm r}^2}\left((u + \delta k_{\rm r})\cosh(\delta k_{\rm r}\tau) + (u\delta k_{\rm r} + 1)\sinh(\delta k_{\rm r}\tau)\right)\right.}
	 \hspace {11cm}\nonumber \\
	 & & \hspace{-9.8cm} +\frac{1}{1+\delta k_{\rm i}^2}\left(\left(-uK_{\rm i} +v -\delta k_{\rm i}\left(\frac{|K|^2 -1}{2}	 + q\right)\right)\cos(\delta k_{\rm i}\tau)\right.
	 \nonumber \\
	 & & \hspace{-7.5cm}\left.\left.\left.\left(\delta k_{\rm i}(uK_{\rm i} -v) - \frac{|K|^2 -1}{2} - q\right)\sin(\delta k_{\rm i}\tau)\right)\right\}\right]
	 \label{b8}
\end{eqnarray}
It is useful to separate out the small contribution to the circular polarisation due to the non-thermal terms (i.e., $u-\xi_{\rm U}$ and $v-\xi_{\rm V}$) from that resulting from $u$ and $v$. Since $\delta k_{\rm r} = -\xi_{\rm U}$, it is convenient to write
\begin{equation}
 	(u\delta k_{\rm r} + 1)\sinh(\delta k_{\rm r}\tau) = (u -\xi_{\rm U})\frac{\sinh(\xi_{\rm U}\tau)}{\xi_{\rm U}} - \frac{u(1-\xi_{\rm U}^2)}{\xi_{\rm U}}\sinh(\xi_{\rm U}\tau).
	\label{b9}
\end{equation}
Furthermore, it is seen from Equations (\ref{b6}) and (\ref{b7}) that  $\delta k_{\rm i}(|K|^2 - 1)/2 - \delta k_{\rm r} K_{\rm i}= \xi_{\rm V}$. With $\delta k_{\rm i} = \hat{\xi}_{\rm U}$ one can then write
\begin{eqnarray}
	\lefteqn{\left(\delta k_{\rm i}(uK_{\rm i} -v) - \frac{|K|^2 -1}{2} - q\right)\sin(\delta k_{\rm i}\tau) =} \hspace{15cm}\nonumber\\
	\left(K_{\rm i}(\xi_{\rm U}-u) + v - \xi_{\rm V} -q\hat{\xi}_{\rm U}\right)\frac{\sin(\hat{\xi}_{\rm U}\tau) }{\hat{\xi}_{\rm U}} + \frac{(1+\hat{\xi}_{\rm U}^2)(uK_{\rm i}-v)}{\hat{\xi}_{\rm U}}\sin(\hat{\xi}_{\rm U}\tau). \nonumber \\
	\label{b10}
\end{eqnarray}
With the use of Equations (\ref{b9}) and (\ref{b10}), the circular polarisation can be expressed as
\begin{eqnarray}
	\lefteqn{V = S\left[K_{\rm i}\left\{\frac{u-\xi_{\rm U}}{1-\xi_{\rm U}^2}\left(1- \exp(-\tau)\left(\cosh(\xi_{\rm U}\tau) + \frac{\sinh(\xi_{\rm U}\tau)}{\xi_{\rm U}}\right)\right) + \exp(-\tau)\frac{u\sinh(\xi_{\rm U}\tau)}{\xi_{\rm U}}\right\}\right.}
	\hspace{15cm}\nonumber \\
	& &\hspace{-14cm}+\frac{K_{\rm i}(\xi_{\rm U}-u) + v -\xi_{\rm V} - q\hat{\xi}_{\rm U}}{1+\hat{\xi}_{\rm U}^2}\left(1-\exp(-\tau)\left(\cos(\hat{\xi}_{\rm U}\tau) + \frac{\sin(\hat{\xi}_{\rm U}\tau)}{\hat{\xi}_{\rm U}}\right)\right)\nonumber \\
	& &\hspace{-12cm}\left. -\exp(-\tau)\frac{(uK_{\rm i} -v)\sin(\hat{\xi}_{\rm U}\tau)}{\hat{\xi}_{\rm U}}\right].	
	\label{b11}	
\end{eqnarray}

\section{Transfer coefficients for a synchrotron source}
\begin{figure}
\epsscale{0.80}
\plotone{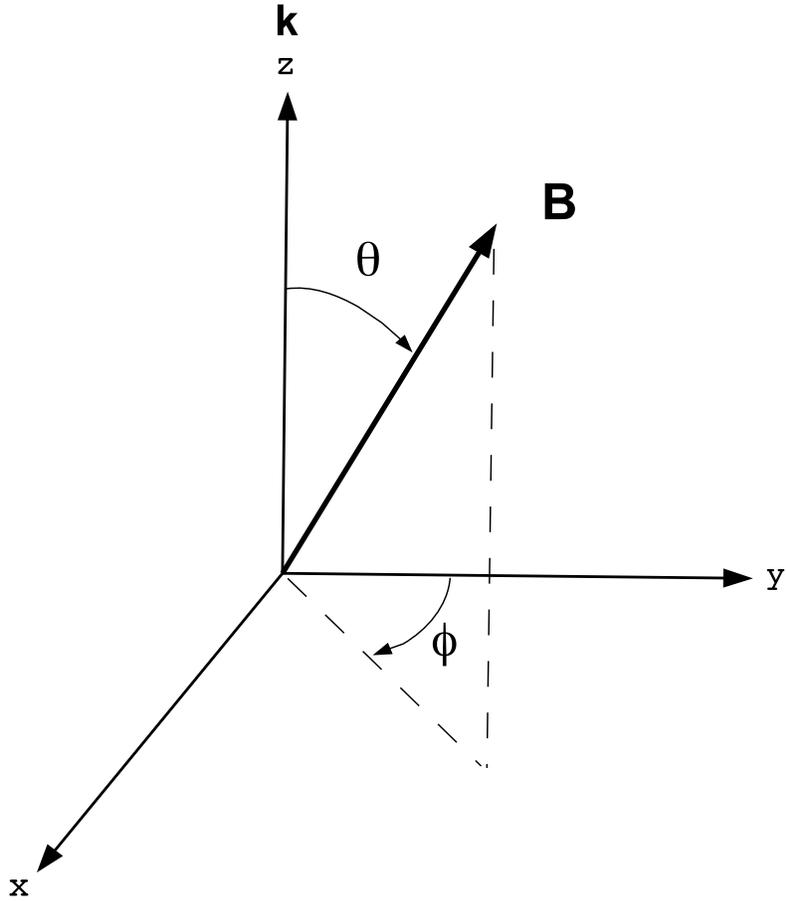}
\caption{The coordinate system used for the transfer equation. The ray propagates along the z-axis and the magnetic field direction is specified by the polar-angle $\theta$ and azimuthal-angle $\phi$.
\label{fig1}} 
\end{figure}

The transfer coefficients for a synchrotron source depend on the energy distribution of the radiating particles. \cite{j/o77} have given the expressions for a power-law distribution (${\rm d}N/{\rm d}\gamma \propto \gamma^{-p}$ for $\gamma > \gamma_{\rm min}$ and $p>2$). The resulting optically thin spectral flux is then $F(\nu) \propto \nu^{-\alpha}$, where $\alpha = (p-1)/2$. The magnitude of the transfer coefficients depends on $p$. However, for realistic values of $p$ this dependence is quite weak \citep[for details, see][]{j/o77}. Hence, neglecting factors of order unity one can write
\begin{eqnarray}
	u &\lsim& \xi_{\rm U}\,\lsim\, 1 \nonumber \\
	|v| &\lsim& |\xi_{\rm V}|\,\lsim \left(\frac{\nu_{\rm B}}{\nu}\right)^{1/2} |\cot \theta|, 
	\hspace{1cm}\mbox {where sign}(v,\xi_{\rm V}) = -\mbox{sign} \cot \theta\nonumber \\
	\hat{\xi}_{\rm U} &\approx& -\left\{\left(\frac{\nu}{\gamma_{\rm min}^2 \nu_{\rm B}}\right)^{(p-2)/
	2} -1\right\}/(p-2)  \nonumber \\
	\hat{\xi}_{\rm V} &\approx& \left(\frac{\nu}{\gamma_{\rm min}^2 \nu_{\rm B}}\right)^{p/2} 	 		\frac{\ln \gamma_{\rm min}}{\gamma_{\rm min}} \cot \theta, 
	\label{c1}	 
\end{eqnarray}
where, $\nu_{\rm B}$ is the cyclotron frequency and $\theta$ is the angle between the line of sight and the magnetic field direction (see Figure \ref{fig1}).

Let $\nu_{\rm abs}$ be the frequency where $F(\nu)$ peaks. The corresponding brightness temperature $T_{\rm b} \propto \gamma_{\rm abs} \approx (\nu_{\rm abs}/\nu_{\rm B})^{1/2}$. Standard synchrotron theory shows that the brightness temperature is quite insensitive to variations in the source parameters \citep[e.g.,][]{b/k17}, 
\begin{equation}
	\gamma_{\rm abs} \propto F(\nu_{\rm abs})^{1/(2p+13)} (U_{\rm e}/U_{\rm B})^{2/(2p+13)}
	\gamma_{\rm min}^{2(p-2)/(2p+13)}, 
	\label{c2}
\end{equation}
where $U_{\rm e}$ and $U_{\rm B}$ are the energy densities in electrons/positrons and magnetic fields, respectively.  For compact radio sources the expected brightness temperature corresponds to $\gamma_{\rm abs} \approx 10^2$. 

The presence of electron-positron pairs effects the polarisation due to the different charge dependences of the transfer coefficients; $u,\,\xi_{\rm U}$, and $\hat{\xi}_{\rm U}$ are unaffected, while the sign of $v,\,\xi_{\rm V}$, and $\hat{\xi}_{\rm V}$ is determined by the charge of the plasma particles. Let $n_{\rm p}$ denote the density of pairs and $n_{\rm exc}$ the excess density of electrons (or positrons). For simplicity, assume that the electrons and positrons have the same energy distributions. This causes the values of $v,\,\xi_{\rm V}$, and $\hat{\xi}_{\rm V}$ to decrease by a factor $n_{\rm exc}/(n_{\rm exc} + 2n_{\rm p})$. 

Including the approximation $\nu_{\rm B} \approx \nu_{\rm abs}/\gamma_{\rm abs}^2 \approx 10^{-4}\nu_{\rm abs}$, Equation (\ref{c1}) can be rewritten as
\begin{eqnarray}
	u &\lsim& \xi_{\rm U}\,\lsim\, 1 \nonumber \\
	|v| &\lsim& |\xi_{\rm V}|\,\lsim\,10^{-2}\left(\frac{\nu_{\rm abs}}{\nu}\right)^{1/2}
	\frac{n_{\rm exc}}{n_{\rm exc} + 2n_{\rm p}} |\cot \theta|,
	\hspace{1cm}\mbox {where sign}(v,\xi_{\rm V}) = -\mbox{sign} \cot \theta \nonumber \\
	\hat{\xi}_{\rm U} &\approx& -\left\{\left(\frac{10^2}{\gamma_{\rm min}}\right)^{p-2}\left(\frac{\nu}
	{\nu_{\rm abs}}\right)^{(p-2)/2} -1\right\}/(p-2) \nonumber \\
	\hat{\xi}_{\rm V} &\approx& 10^{2p}\left(\frac{\nu}{\nu_{\rm abs}}\right)^{p/2}
	 \frac{\ln \gamma_{\rm min}}{\gamma_{\rm min}^{1+p}} \frac{n_{\rm exc}}{n_{\rm exc} + 
	 2n_{\rm p}} \cot \theta. 
	\label{c3}	 
\end{eqnarray}
It may be noted from Equation(\ref{c3}) that the limit of nearly circular characteristic waves (i.e., $|\hat{\xi}_{\rm V}| \gg |\hat{\xi}_{\rm U}|$) implies $\gamma_{\rm min} \ll \gamma_{\rm abs} \approx 10^2$ so that  
\begin{equation}
	\frac{\hat{\xi}_{\rm U}}{\hat{\xi}_{\rm V}} \approx -\frac{10^{-4}}{p-2}\frac{\nu_{\rm abs}}
	{\nu} \frac{\gamma_{\rm min}^3}{\ln \gamma_{\rm min}}\left(1+\frac{2n_{\rm p}}{n_{\rm exc}}\right) \cot ^{-1} \theta.
	\label{c4}
\end{equation}

\clearpage

\end{document}